 \documentclass[12pt]{article}
 \usepackage{amsmath,amssymb}
 \usepackage{booktabs}
 \usepackage{graphicx}
 \usepackage[most]{tcolorbox}
 \usepackage[margin=1in]{geometry}
 \newenvironment{sciabstract}{\begin{quote} }{\end{quote}}

\title{What Quantitative Risk Modellers Can Learn from Durkheim's Study of Suicide}

\author{Mahmood Alaghmandan,
\
\\
\normalsize{\it Manager, Model Validation and Risk Management}\\
\normalsize{Farm Credit Canada,}\\
\normalsize{ 100 Queen St, unit 1460, Ottawa, ON K1P 1J9 Canada}\\
\small{{\bf Email:} \texttt{mahmood.alaghmandan@fcc.ca}}
}

\date{\today}

\begin{document}

\baselineskip24pt

\maketitle

\begin{center}
\begin{minipage}{0.75\textwidth}
\centering
\itshape
To M! To those who knew you, you were---and will always be---so much more than a row in a suicide dataset.\\
\end{minipage}
\end{center}

\vskip2cm

\begin{sciabstract} {\bf Abstract:} \'{E}mile Durkheim's \emph{Suicide: A Study in Sociology} (1897) predates much of the statistical machinery that quantitative modellers now take for granted. Yet, working with sparse and imperfect observational data, Durkheim repeatedly arrives at practices that remain remarkably relevant to modern modelling. This paper revisits \emph{Suicide} from the perspective of quantitative risk modelling, not for its substantive conclusions, but for the reasoning by which Durkheim reached them. His approach illustrates how careful definition, common sense, logical investigation, scepticism toward convenient explanations, and close attention to what the data can and cannot support—all of which must precede, and can often substitute for, statistical sophistication.
The broader lesson is simple: good modelling begins not with technique, but with understanding the problem, interrogating the evidence, and reasoning carefully about what it can actually tell us. \end{sciabstract}

\section{Introduction}

In 1897, \'{E}mile Durkheim, a French sociologist, published Suicide: A Study in
Sociology\footnote{Le Suicide: \'{E}tude de sociologie. \cite{durk}}. The book remains a landmark work in
the study of suicide, its underlying causes, and its social context. 
Although it was only the second major empirical study of suicide as a social phenomenon\footnote{Tom\'{a}\v{s} Masaryk published Suicide
as a Social Mass Phenomenon of Modern Civilization in 1881. \cite{masa}}, its methodology was unprecedented within sociology.

In this book, which remains widely read and frequently cited on the subject to this day,
Durkheim uses the data available to him to systematically examine a series of propositions
concerning the social determinants of suicide, and from these carefully draws conclusions on
its nature. The analytical approach he employs is both meticulous and remarkably thorough.
Throughout his investigation, he demonstrates a deep awareness of what the data can
legitimately support, while remaining equally attentive to their limitations. His careful
treatment of observational data, his reluctance to draw conclusions that exceed the evidence,
and his systematic consideration of alternative explanations embody principles that should
guide any form of inductive reasoning.

While I am not a sociologist, I found Durkheim's meticulous and rigorous scientific investigation in this
classic work fascinating. As a quantitative analyst with extensive training in the
mathematical sciences and their applications, I was particularly struck by the care with which
he formulated and tested his hypotheses. I believe that quantitative risk modellers have much
to learn from his approach to data, the conclusions that can legitimately be drawn from it,
and the limitations that must be acknowledged. In this paper, I share my observations on
Durkheim's methodology from the perspective of a quantitative risk modeller.

Two clarifications are in order before proceeding. The first concerns scope. The purpose of this article is not to assess the accuracy or present-day validity of Durkheim's substantive claims about suicide. Several have been contested, refined, or overturned in the intervening century, while the social conditions on which they were premised have changed substantially\footnote{The literature on this point is extensive; a modeller's summary would be that the dependent variable, the covariates, and the measurement process have all drifted. This is itself a useful lesson: a model estimated on nineteenth-century European administrative data has no claim on twenty-first-century populations, however elegant its construction.}. What remains durable in \emph{Suicide} is not necessarily its answers, but its manner of arriving at them. Accordingly, this article is about Durkheim's method, and about suicide only incidentally.

The second concerns the conditions under which he worked, which are precisely what make the book instructive. Durkheim was working with sparse, imperfect administrative data and without most of the statistical and computational machinery available to a modern modeller. He therefore had little choice but to carry his investigation through reasoning: making assumptions explicit, confronting alternative explanations, and asking what observable patterns each proposed mechanism should produce.

This is, I would argue, precisely the discipline that modern quantitative practice has made easier to neglect. Statistical sophistication can greatly improve an analysis, but it can also make it possible to obtain an answer without first reasoning carefully about why that answer should be believed. The growing use of artificial intelligence makes this distinction more important still: increasingly sophisticated analyses can now be produced with increasingly little effort, while the responsibility for understanding what is being modelled, interrogating the evidence, and judging what conclusions it can support remains with the modeller. The remainder of this paper examines several practices in Durkheim's investigation that remain instructive for model development, validation, and model risk management.\footnote{Throughout, I use ``modeller'' to cover both the developer and the validator. The practices below are the developer's responsibility to apply and the validator's responsibility to test.}

\begin{tcolorbox}
\textbf{What this paper offers.}
This is not a new statistical methodology, but something more elementary—and perhaps more easily neglected: a set of practical modelling disciplines grounded in common sense, logical investigation, and careful attention to what the data can---and cannot---tell us.
\end{tcolorbox}

\noindent\textbf{Author's Note.}
The views expressed in this paper are solely those of the author and should not be attributed to the author's current or former employer.

\section{Definition Before Measurement}\label{s:definition}

Durkheim opens the book not with an analysis or a hypothesis, but with a definition, the definition of suicide, and he spends a full chapter on it.
He is dissatisfied with the common-sense understanding of his subject because it smuggles in
intent, which is unobservable, and because it fails to draw a stable boundary around the class of
events to be counted. He settles instead on a definition framed entirely in terms of the act and
the actor's knowledge of its consequences to address all the concerns he raised earlier. He then examines the boundary cases his definition admits or excludes, and accepts the consequences of where he has drawn the line.

The methodological point is not which definition he chose. It is that he chose one, stated it, tested it against edge cases, and held to it for four hundred pages—all \emph{before} looking at a single table.

The parallel in quantitative risk modelling is immediate and, in my experience, the single most
common source of avoidable model error. Consider default. A modeller building a probability of default model is estimating the probability of an event—the default of an obligor—whose definition is a modelling choice, not a fact of nature: ninety days past due, or ninety days past due excluding technical arrears; unlikeness to pay as assessed by whom, and recorded where; does a restructuring count, and if so from which
date; is the cure period thirty days or twelve months; does a facility that defaults, cures, and
re-defaults contribute one observation or two. Each of these choices changes the estimated default
rate, sometimes by a factor that dwarfs the effect of the model form. 

The same applies to loss
given default (is the recovery discounted, at what rate, over what workout horizon, and are
indirect costs allocated), to exposure at default, to the definition of a ``stressed'' period in a
market risk model, and to the target variable in essentially every credit model I have reviewed.

Durkheim's discipline translates into three requirements:

\begin{enumerate}
\item \textbf{The definition of the modelled event precedes the extraction of the data.} If the
definition is arrived at after the data have been inspected, the modeller cannot distinguish a
definition chosen for conceptual soundness from one chosen because it produced a defensible
default rate. This is not an accusation of bad faith; it is an observation about the
irreversibility of having seen the answer.

\item \textbf{The definition is tested against boundary cases before use.} Durkheim asks whether a
soldier's sacrifice, a martyr's death, and a refusal of medical treatment fall inside his
definition, and answers each explicitly. The modelling analogue is a documented set of adjudicated
edge cases---the technically-past-due-but-performing account, the fraud-driven write-off, the
facility transferred between entities mid-workout---resolved on paper and then applied
mechanically.

\item \textbf{The definition used in estimation matches the definition used in application.} A
model estimated on an accounting default definition and deployed against a regulatory one is
mis-specified in a way that no amount of backtesting on the estimation sample will reveal. This
mismatch is a recurring finding in validation and is almost always traceable to the definition
having been treated as a data-extraction detail rather than as a modelling decision.
\end{enumerate}

A useful discipline, borrowed directly from Durkheim's chapter structure, is to require that the
first substantive section of every model development document define the modelled quantity in
terms that a reader outside the modelling team could apply to an individual record without further
guidance. If that cannot be done, the model is not yet ready to be built.

\begin{tcolorbox}

{\bf Define before you model.}
The first substantive section of every model development document should answer one question:

\begin{center}
\emph{What, exactly, is being modelled?}
\end{center}

The definition should be precise enough that a reader outside the modelling team could apply it to an individual record without further guidance. {If the modelled quantity cannot yet be defined, the model is not yet ready to be built.}

\end{tcolorbox}

\section{Data Sanity Checks and Negative Controls}

Durkheim's data were official mortality statistics compiled by state bureaux for administrative
purposes. He did not treat them as ground truth. He asks, in effect, how a death comes to be
recorded in the category he is studying: who makes the determination, under what incentives, with
what evidence, and with what variation across jurisdictions and over time. He is explicit that
some deaths are misclassified, that the direction of misclassification is not random, and that
jurisdictions differ in their recording practices in ways correlated with the very social
characteristics he is investigating. Where he relies on a figure, he first asks what process
produced it.

He is at his most rigorous, however, in what modern practice would call \textbf{negative controls}. Before advancing any social explanation, he devotes a substantial portion of the book to systematically eliminating the non-social explanations that were prevalent at the time, including imitation, race, climate, and temperature factors. In each case, he begins by clearly defining the proposed explanation and assessing whether it can be meaningfully captured as a risk driver. In some instances, such as imitation and race, this proves impossible. 

Where empirical evaluation is feasible, however, he does not merely assert that the explanation is inadequate. Rather, he derives the patterns that would be expected if the explanation were correct and then demonstrates that the observed data fail to exhibit those patterns. The temperature hypothesis provides the clearest example. If a drop in temperature were the driving factor, as commonly believed, the highest suicide rates should occur during the coldest months of the year and in colder regions.  They do not. Instead, suicide rates increase with the length of the day and with the intensity of collective activity.  Moreover, this relationship persists even when temperature is held approximately constant across regions. The candidate explanation is therefore rejected not because it appears implausible, but because it fails an empirical prediction.

For the risk modeller, these sections of \emph{Suicide} are a description of good data governance and
challenger-hypothesis testing, written a long time before either had a name.

\paragraph{Provenance before profiling.} Every risk model is estimated on data assembled for a
different purpose. The modeller who does not know which system a
field comes from, when its definition last changed, and what a null in that field means, is not in a
position to interpret the estimated coefficients. 

The recurring failures are familiar: a field whose
meaning changed at a system migration and now carries two populations under one name; a default flag
that is set at the account level in one legacy portfolio and at the obligor level in another; a
collateral value that is sometimes an appraisal, sometimes an indexed appraisal, and sometimes a
placeholder. None of these are detectable from the fitted model. All are detectable by asking
Durkheim's question: what process produced this data input?

\paragraph{Sanity checks with a stated expectation.} 
A sanity check is meaningful only if the criteria for sanity are clearly defined in advance.
 For example, aggregate default rates should be reconciled to the
regulatory and financial reporting figures for the same period, and the difference explained, not
merely noted. Exposure totals should tie to the balance sheet. Rating migration matrices should be dominated
by the diagonal. Where a check fails, the failure is a finding about the data, not a nuisance to be
filtered out.

\paragraph{Negative controls proper.} Perhaps the most valuable—yet least practised—of Durkheim's techniques is deliberately testing a model against an outcome or a population for which the hypothesized effect should be absent, and confirming that it is indeed absent. For example, if a behavioural score discriminates well in the development sample, does it also appear to discriminate against an outcome that it cannot plausibly influence?

Such analyses can be particularly valuable when business experts encourage a development team to incorporate risk drivers simply because they proved useful in the past or in a different environment. A modeller owes stakeholders more than either categorically rejecting such requests or reluctantly accepting them. Instead, the proposed risk drivers should be subjected to careful empirical evaluation. As Durkheim demonstrated, this requires a deeper analysis than a simple hypothesis test or the reporting of a p-value. By deriving the empirical implications of the proposed explanation and confronting them with the data, the modeller can provide persuasive evidence to stakeholders about whether the proposed risk driver truly belongs in the model.

\begin{tcolorbox}

{\bf Good modelling begins before estimation and continues beyond statistical significance.}
Durkheim's approach suggests three practical disciplines:

\begin{enumerate}

\item \textbf{Know where the data came from.} Understand how each input was generated, what it means, and whether its definition has changed. A fitted model cannot reveal broken provenance.

\item \textbf{Know what ``sane'' should look like.} Define expectations before checking the data, and reconcile important quantities to independent benchmarks. A failed sanity check is a finding, not a nuisance.

\item \textbf{Test where the effect should disappear.} Challenge a proposed risk driver against outcomes or populations where its hypothesized effect should be absent. Statistical significance alone is not enough; a credible explanation should survive attempts to falsify it.

\end{enumerate}

{Do not merely ask whether the data fit the hypothesis. Ask what else should be true---and what should not be true---if the hypothesis is right.}

\end{tcolorbox}

\section{Outliers Carry Information}

One of the interesting findings in \emph{Suicide} is the relationship between religion and suicide rate: Protestant populations exhibited markedly higher rates than Catholic ones, a pattern Durkheim observed both across countries and across regions within countries. Durkheim explained this phenomenon through a careful comparison of the two branches of Christianity and their expectations of believers, specifically, how Catholicism prescribes the faith for its followers and binds them into a common ritual life, while Protestantism grants the believer greater latitude for individual interpretation and free inquiry. His account—that the two confessions differed in the degree to which they bound the individual into a prescribed collective life—is the origin of his concept of social integration.

What matters here is not that finding but the one that follows it. On Durkheim's data, Jewish populations showed suicide rates lower than both Protestants and Catholics, despite being a minority under sustained social hostility. The puzzle was sharper still: Jews were also highly educated and deeply engaged in free inquiry, a trait Durkheim had identified elsewhere as a driver of \emph{elevated} suicide risk, since it was his explanation for the Protestant rate itself. By his own covariate, Jewish communities should have ranked near the top; instead they ranked at the bottom. This is an outlier in the strict sense—a point the fitted relationship does not accommodate.

Durkheim does not exclude this observation from his sample, winsorize it, or simply note that it falls outside the general pattern and move on. Instead, he treats it as an informative observation, devoting an entire chapter to understanding it. It is this observation that compels him to refine his earlier explanation: what is protective against suicide, in his account, is neither doctrine, free inquiry, nor level of education, but rather the density of the social obligations that bind a community together. A minority living under external pressure can exhibit such cohesion to an exceptional degree, and this is what placed Jewish communities well above Catholics and farthest from Protestants. The outlier does not merely survive the analysis—it changes the model.

The financial modelling analogue is uncomfortable, because the standard treatment of an outlier in
a development pipeline is removal or capping, and the standard justification is that the observation
would otherwise exert undue influence on the fit. That justification is sometimes correct---a
keying error, a duplicated record, a test account---and Durkheim would have agreed, since he took
care to establish that a figure was real before reasoning from it. But the justification is
frequently applied to observations that are entirely real and merely inconvenient.

Consider the segments that generate them: the counterparty whose loss rate is an order of magnitude
above its rating cohort; the single vintage that fails to season; the region whose delinquency did
not respond to a rate shock that moved every other region; the obligor group whose recoveries
exceeded exposure. Each of these is a point at which the model's implied mechanism is contradicted
by the portfolio. Removing it improves the fit statistics and destroys the information. In a stress
testing context the cost is direct: the observations most likely to be trimmed as outliers are, by
construction, drawn from the tail that the stress model exists to describe.

\begin{tcolorbox}

{\bf An unusual observation should trigger investigation, not automatic exclusion.}
For every observation considered for exclusion, a modeller should ask---and document---three questions:

\begin{enumerate}

\item \textbf{Is it real?} Verify the observation against the source data. Distance from the mean is a symptom, not a diagnosis.

\item \textbf{If real, what does it contradict?} Identify the model assumption it violates. If it violates none, it may simply be an extreme but valid observation.

\item \textbf{Can the model accommodate it?} Consider segmentation, additional covariates, interactions, or a different functional form before resorting to exclusion.

\end{enumerate}

{If exclusion is ultimately justified, document why---and be able to show what the model would have concluded had the observation been retained.}

\end{tcolorbox}

\section{Depth of Investigation: Confounders and Adjustment Factors}

Durkheim's treatment of the relationship between marital status and suicide rates is, to my mind, the most impressive piece of quantitative reasoning in Suicide and one that ought to be required reading for anyone who builds models from observational data.

The raw comparison is straightforward: married and unmarried populations exhibit different suicide rates, with married individuals appearing to die by suicide at a substantially higher rate. At first glance, this seems to support the commonly held belief that marriage increases the risk of suicide because of the burdens and stresses of family life.

Yet Durkheim dismisses this conclusion immediately. The raw comparison is fundamentally misleading because married people are, on average, older than unmarried people, and suicide rates vary substantially with age. Any observed difference between the two groups therefore confounds the effect of marriage with the effect of age. In an era before regression analysis, there was no automatic statistical remedy for this problem.

His response is to do the work by hand. He reorganizes the data into age bands and compares married and unmarried individuals within each band, constructing what he calls a \emph{coefficient of preservation}: the ratio of the suicide rate among the unmarried to that among the married of the same age. In modern terminology, this is a stratified, age-standardised relative risk, computed in 1897 without the concept or its accompanying statistical framework being available to him.

The stratification is not merely a technical refinement; it fundamentally changes the conclusion. In his tables, young married men exhibit lower suicide rates than their unmarried contemporaries, reversing the na\"{i}ve inference drawn from the aggregate data and forcing the reader to seek a more careful explanation of the underlying mechanism.

He then takes the analysis one step further by asking what the coefficient is actually measuring. Married and unmarried individuals differ in more than just age, and Durkheim argues that much of the protective effect initially attributed to marriage is, in fact, associated with the presence of children rather than marriage itself.

There are two lessons here, both of which I wish were more commonly applied in our modelling practices, well over 130 years after Durkheim arrived at them largely through careful reasoning and common sense.

\begin{enumerate}
\item{{\bf Pursue confounding as far as the data allow:}
Where the data permit stratification, stratify; where they permit the inclusion of a control variable, include it. The analogue in credit modelling is immediate. A rating grade may correlate with facility size, facility size with industry, industry with collateral type, and each of these with the vintage in which the facility was originated. An observed difference in performance between two rating grades, products, or channels may therefore reflect some combination of these factors rather than the characteristic under investigation.

The failure mode is familiar to any validator: a difference in performance across segments is presented as evidence of a genuine segment effect, even though the segments differ systematically in other relevant characteristics, such as seasoning. The objective should therefore not be merely to establish that the groups differ, and therefore simply model them independently, but to ensure that the model can meaningfully explain those differences and distinguish the effect of the characteristic of interest from the effects of the factors with which it is confounded.}

\item{ \textbf{When the data do not permit adequate control, make the adjustment explicit:}
Durkheim did not always have the joint distribution required to isolate an effect cleanly. When the necessary data were unavailable, he constructed a correction from the information he did have and stated explicitly that an adjustment had been made. The reader could therefore see the correction, assess its rationale, and, if unconvinced, discount the resulting conclusion accordingly.

There is a direct analogue in econometric modelling. Suppose an explanatory variable is known to be affected by a factor that is not observed in the modelling dataset, but information about that factor is available from an external study, an aggregate dataset, or a related population. Rather than ignoring the missing factor, the modeller may construct a reasonable adjustment or calibration based on that external evidence. The resulting estimate is not equivalent to what could have been obtained had the missing variable been observed directly, and it should not be presented as such. The adjustment, its empirical basis, and the assumptions required for its use should be reported transparently.

Where the adjustment factor itself is uncertain, sensitivity analysis becomes particularly important. If, for example, the best available evidence suggests an adjustment of 10\%, the analysis need not rest entirely on the assumption that 10\% is the correct value. The modeller can repeat the analysis under plausible alternatives---say 5\%, 10\%, and 15\%---and examine whether the substantive conclusion survives. If it does, the conclusion is more credible despite the data limitation; if it does not, that instability is itself an important result and should be reported.

This is an important middle ground that modelling practice can easily overlook. The absence of ideal data does not necessarily require either ignoring a known deficiency or abandoning the analysis altogether. A reasonable adjustment, explicitly identified and supported by the available evidence, may be preferable to both. But the distinction between what is observed and what is assumed must remain visible, and where the assumption is consequential, its uncertainty should be carried through the analysis.}

\end{enumerate}

\begin{tcolorbox}

\textbf{Understanding before technique.}
None of these observations would have revealed themselves to Durkheim had he not followed two fundamental principles:

\begin{itemize}

\item[(i)] \textbf{Understand before you model.} Know what is being modelled, which risk factors matter, why they matter, and through what mechanisms they affect the outcome.

\item[(ii)] \textbf{Statistics cannot substitute for understanding.} They reveal associations; they do not tell the modeller which comparisons are meaningful, which variables are confounders, or which adjustments are defensible.

\end{itemize}

\begin{center}
\emph{A modeller should be able to explain the data, not merely reproduce them mathematically.}
\end{center}

\end{tcolorbox}

\section{Resolution: Using the Data at the Finest Level Available}

A striking feature of \emph{Suicide}, and one easy to overlook, is how hard Durkheim works the
granularity of his sources. He has country-level figures for much of Europe, and he uses them. But
where French administrative data are available at the level of the \emph{d\'{e}partement}, he drops
to that level, and the additional resolution does real analytical work: it lets him separate a
national explanation from a regional one, observe a gradient rather than a contrast, and test whether
a pattern claimed at one level of aggregation survives at a finer one.

Two things follow, and the second is the constraint on the first.

\paragraph{Aggregate first, analyse second, is backwards.} A pattern visible at the country level may
be an artefact of composition---the ecological fallacy, of which \emph{Suicide} is incidentally one
of the most-discussed examples in the methodological literature. Durkheim's use of departmental data
is the partial defence available to him: a relationship that holds at country level, at regional
level, and within regions is far more credible than one that holds only after aggregation. 

Consider how physical climate risk is measured across different hazards. Aggregation could be  a serious mistake here, particularly for hazards whose impact varies sharply across small geographic distances. Pluvial flooding is a clear case: two properties a few hundred meters apart can face entirely different risk profiles depending on whether one sits at the bottom of a slope or the top of it. A regional or national average flood-loss figure obscures this; it blends locations that will flood in almost any storm with locations that will almost never flood, producing a number that describes neither.

\paragraph{Resolution is a property of the data, not of the ambition.} Durkheim goes to the \emph{département} level where the data permit and does not pretend to it where they do not. Modern modelling often violates this discipline in both directions. At one extreme, granular data are aggregated to whatever level makes modelling easiest, sacrificing information the data could support. 

At the other extreme, models are built at a granularity the data cannot sustain. A portfolio segmentation with forty cells and eleven defaults is not fine-grained; it is a coarse dataset spread across too many parameters. What matters is not the total number of records, but the number of \emph{events} supporting each estimated parameter---and, for time-series models, the number of independent macroeconomic cycles in the estimation sample, which is very often one.

\begin{tcolorbox}

\textbf{Use the resolution the data earn.}
Two principles should govern the choice of modelling granularity:

\begin{itemize}

\item[(i)] \textbf{Do not model finer than the data support.} Push to the finest resolution the data genuinely sustain---and stop there.

\item[(ii)] \textbf{Do not aggregate away information without investigating it.} If the data contain meaningful variation at a finer level, examine it before averaging it away. Aggregation may ultimately be necessary, but it should be a conclusion of the analysis, not its starting point.

\end{itemize}

\begin{center}
\emph{Use all the resolution the data support, but no more.}
\end{center}

\end{tcolorbox}

\section{Causality versus Correlation: The Imitation Hypothesis}

The chapter of \emph{Suicide} that best repays a modeller's attention is Durkheim's treatment of imitation. Gabriel Tarde had assigned imitation a central role in the propagation of social phenomena, and suicide appeared an especially plausible case: individual acts could demonstrably provoke similar acts in others, and apparent clusters were readily interpreted as evidence of contagion. The hypothesis was prominent, intuitively appealing, and supported by seemingly suggestive observations. Durkheim rejects it as an explanation of the social suicide rate, and the manner of the rejection is the point.

He begins where a modeller should begin and where, in practice, few do: with a clear definition (see Section~\ref{s:definition} on its importance). He observes that ``imitation'' is used loosely to describe several distinct processes---the emergence of a shared collective sentiment, conformity to customs and social authority, and the automatic reproduction of an observed act---despite their entirely different mechanisms. Only the last is imitation in the sense relevant to the hypothesis being tested. His first move is therefore to make the hypothesis testable by defining precisely what would count as imitation. Without that step, as Durkheim observes, a word can easily be mistaken for an explanation.

This was the observation, recorded on the reverse of an old napkin, that prompted this section: \emph{imitation is hard to define}. An undefined mechanism cannot meaningfully be confirmed or refuted; it can only be illustrated.

Having sharpened the definition, he asks what a genuine contagion process would leave behind in the
data. If cases propagate by imitation, they must propagate \emph{from} somewhere: there should be
foci, and rates should decay with distance from those foci, and the pattern should be indifferent
to social boundaries that do not impede the transmission of information. He then examines the
geographic distribution and finds nothing of the kind. Rates cluster, but they cluster along
regional and social lines, without radial decay from centres, and the clustering is explained by the
same social characteristics that explain the level. Cases co-occur because the people involved share
an environment, not because one caused the next.

This is a complete piece of causal reasoning conducted without any of the tools we would now use: a
mechanism stated precisely enough to generate a testable implication, an observable signature derived
from it, a confrontation with data, and the elimination of the hypothesis on the failure of the
signature rather than on the analyst's prior.

The application to quantitative risk modelling hardly needs spelling out, but it is worth being
concrete about where the failure occurs. Risk models are full of relationships that are treated as
causal because they are convenient to project. A macroeconomic variable enters a stress model because
it correlates with historical losses and because the regulator publishes a path for it. A behavioural
variable enters a scorecard because it lifts discrimination. In neither case is the mechanism
typically stated, and in neither case is the alternative---that both series respond to a third
factor, or that the relationship reflects the composition of the sample---typically eliminated.

The consequence is not visible in normal conditions, because a variable that co-moves for the wrong
reason could still co-move for a while. It becomes visible exactly when the model is relied upon: under stress, when
the relationship between the driver and the loss is asked to hold outside the range in which it was
observed, and when the third factor that produced the historical co-movement is no longer operating
in the same way. A model that cannot state its mechanism cannot state the conditions under which its
mechanism fails, and therefore cannot be assigned a range of validity.

\begin{tcolorbox}

\textbf{Make every risk driver explain itself.}
For every driver retained in a model, the developer should document three things:

\begin{itemize}

\item[(i)] \textbf{State the mechanism.} Explain how and why the driver is expected to affect the outcome, in terms of borrower or counterparty behaviour rather than statistical fit alone.

\item[(ii)] \textbf{State an alternative explanation.} Identify at least one plausible reason for the observed relationship other than the proposed mechanism---particularly a common cause---and consider what limitations it would impose on the use of the driver.

\item[(iii)] \textbf{Try to distinguish between them.} Where the data permit, investigate the competing explanations quantitatively; where they do not, use qualitative evidence and subject-matter expertise rather than leaving the question unexamined.

\end{itemize}

\begin{center}
\emph{A driver that cannot be explained cannot be assigned a reliable range of validity.}
\end{center}

\end{tcolorbox}

\section{Concluding Remarks}

It would be a poor reading of this paper to conclude that quantitative modellers should emulate a nineteenth-century sociologist because his conclusions were right. Several were not, and the ecological inference on which much of his argument rests is now a textbook illustration of a methodological hazard. That is not the point. 

Durkheim himself was explicit that his conclusions were provisional and bounded by the data available to him.
The point is that the constraints under which he worked imposed a discipline that abundance has eroded. He had no algorithm to choose his variables, so he had to argue for each one. He had no significance test to substitute for judgement about whether a difference was meaningful, so he had to reason about the process that generated the numbers. He could not fit a flexible functional form and report an out-of-sample error, so he had to state what his hypothesis predicted and examine whether the evidence behaved accordingly. Modern modelling has removed many of these constraints. In doing so, it has also made the reasoning they once enforced optional.

A reader who approaches \emph{Suicide} from this perspective will find far more than these few sections can convey: the patience with which Durkheim establishes when populations are sufficiently comparable to support analysis; the care with which he examines their heterogeneity before permitting himself to compare them; and, underlying both, the modelling philosophy by which he justifies treating an individual act as a social phenomenon susceptible to measurement at all. He builds that philosophy explicitly, defends it against objections, and then holds himself to it throughout the book.

The enduring lesson is therefore not that modern modellers need Durkheim's methods. We have better ones. It is that better methods do not relieve us of the need for the common sense, logical investigation, and disciplined reasoning that were necessary when those methods did not yet exist---a lesson that becomes more, not less, important as artificial intelligence makes sophisticated analysis increasingly easy to produce without requiring an equally sophisticated understanding of what is being analysed.


\end{document}